\pdfoutput=1

\documentclass[lettersize,journal]{IEEEtran}
\usepackage[noadjust]{cite}
\usepackage{ragged2e}
\usepackage{amssymb}
\usepackage{pifont}
\newcommand{\cmark}{\ding{51}}%
\newcommand{\xmark}{\ding{55}}%
\usepackage{amsmath,amsfonts}
\usepackage{algorithmic}
\usepackage{array}
\usepackage[caption=false,font=footnotesize]{subfig}
\usepackage{textcomp}
\usepackage{capt-of}
\usepackage{stfloats}
\usepackage[hyphens]{url}
\usepackage{verbatim}
\usepackage{graphicx}
\usepackage{adjustbox}
\usepackage{tikz}
\usepackage{scalerel}
\usetikzlibrary{svg.path}
\usetikzlibrary{positioning,chains}
\usepackage{multirow}
\usepackage{color, colortbl}
\usepackage{lipsum,xcolor}
\usepackage{graphicx,midfloat,capt-of}
\usepackage{forest}
\usepackage{tikz-qtree}
\usepackage{multicol}
\usepackage{acro}
\usepackage[T1]{fontenc}
\usepackage[utf8]{inputenc} 
\usetikzlibrary{calc,positioning}
\usetikzlibrary{arrows,shapes,positioning,shadows,trees}
\usetikzlibrary{patterns}
\usepackage{svg}
\usepackage{amsmath}
\usepackage{pgfplots}
    \usepgfplotslibrary{
        groupplots,
        units,
    }
    \pgfplotsset{compat=newest}

\usepackage{flafter}

\def\BibTeX{{\rm B\kern-.05em{\sc i\kern-.025em b}\kern-.08em
    T\kern-.1667em\lower.7ex\hbox{E}\kern-.125emX}}
\usepackage{balance}

\usepackage{hyperref}       
\hypersetup{breaklinks=true}

\begin{document}
\pgfplotsset{
    group gap/.initial=0.15cm,
    group size/.initial=2,
    kynes axis/.style={
        plot1/.style={fill=red!20},
        plot2/.style={fill=blue!20},
        ybar=0pt, 
        ymin=0, ymax=11,
    },
    kynes right axis/.style={
            xbar=0pt,
            y axis line style = { opacity = 0 },
            axis x line       = none,
            tickwidth         = 0pt,
            enlarge y limits  = true,
            enlarge x limits  = true,
            ytick distance=1.1,
            ytick=data,
            xtick=\empty,
            legend cell align={left},
            legend style={draw=none,at={(0.1,-0.05)},anchor=north},
            y dir=reverse,
            y=\pgfkeysvalueof{/pgfplots/group size}*\pgfkeysvalueof{/pgf/bar width}+\pgfkeysvalueof{/pgfplots/group gap},
            enlarge x limits={abs=\pgfkeysvalueof{/pgfplots/group size}/2*\pgfkeysvalueof{/pgf/bar width}+\pgfkeysvalueof{/pgfplots/group gap}},
    }
}

\title{Towards a Taxonomy of Industrial Challenges and Enabling Technologies in Industry 4.0}

\author{Roberto Figliè, Riccardo Amadio, Marios Tyrovolas, \IEEEmembership{Student Member, IEEE}, Chrysostomos Stylios, \IEEEmembership{Senior Member, IEEE}, Łukasz Paśko, Dorota Stadnicka, Anna Carreras-Coch, Agustín Zaballos, Joan Navarro, and Daniele Mazzei,

\thanks{This work has been partially funded by Programme Erasmus+, Knowledge Alliances, Application No 621639-EPP-1-2020-1-IT-EPPKA2-KA, PLANET4: Practical Learning of Artificial iNtelligence on the Edge for indusTry 4.0.}
\thanks{Roberto Figliè, Riccardo Amadio, and Daniele Mazzei are with the
Computer Science Department, University of Pisa, Pisa, 56127 Italy (emails:  roberto.figlie@phd.unipi.it, r.amadio@studenti.unipi.it, daniele.mazzei@unipi.it).}
\thanks{Marios Tyrovolas, and Chrysostomos Stylios are with the Department of Informatics and Telecommunications, University of Ioannina, Arta, 47150 Greece. Chrysostomos Stylios is also with Industrial Systems Institute, Athena Research Center, Patras, 26504 Greece (e-mails: tirovolas@kic.uoi.gr, stylios@isi.gr).}
\thanks{Łukasz Paśko, and Dorota Stadnicka are with the Faculty of Mechanical Engineering and Aeronautics, Rzeszów University of Technology, Rzeszów, 35959 Poland (e-mails:  dorota.stadnicka@prz.edu.pl, lpasko@prz.edu.pl).}
\thanks{Anna Carreras-Coch, Agustín Zaballos, and Joan Navarro are with Research Group in Internet Technologies \& Storage, Universitat Ramon Llull, Barcelona, 08022 Spain (e-mails: anna.carreras@salle.url.edu, agustin.zaballos@salle.url.edu,  jnavarro@salle.url.edu).}}

\maketitle

\begin{abstract}
Today, one of the biggest challenges for digital transformation in the Industry 4.0 paradigm is the lack of mutual understanding between the academic and the industrial world. On the one hand, the industry fails to apply new technologies and innovations from scientific research. At the same time, academics struggle to find and focus on real-world applications for their developing technological solutions. Moreover, the increasing complexity of industrial challenges and technologies is widening this hiatus. To reduce this knowledge and communication gap, this article proposes a mixed approach of humanistic and engineering techniques applied to the technological and enterprise fields. The study's results are represented by a taxonomy in which industrial challenges and I4.0-focused technologies are categorized and connected 
through academic and grey literature analysis. This taxonomy also formed the basis for creating a public web platform where industrial practitioners can identify candidate solutions for an industrial challenge. At the same time, from the educational perspective, the learning procedure can be supported since, through this tool, academics can identify real-world scenarios to integrate digital technologies' teaching process.
\end{abstract}


\section{Introduction} \label{introduction}

\IEEEPARstart{T}{he} Fourth Industrial Revolution (also called Industry 4.0 or I4.0) has become the latest flagship in approaching modern technological developments to the new industrial era that humankind is exposed to nowadays \cite{RePEc:eee:proeco:v:204:y:2018:i:c:p:383-394}. This situation has been motivated by several factors \cite{BAI2020107776}: changes in societal patterns (e.g., from physical and social networks to digital interconnection dependencies) and evolution of processes (e.g., from paper-based processes to business digitalization), the broad availability of advanced (also referred to as smart) technologies or pillars (e.g., Artificial Intelligence -AI-, Edge Computing -EC-, Industrial Internet of Things -IIoT-) at affordable costs \cite{russmannIndustryFutureProductivity}, or the dawn of Big Data, among others. Different from the Third Industrial Revolution \cite{greenwoodThirdIndustrialRevolution1999}, that was aimed to use electronics and information technologies to automate production, Industry 4.0 proposes to merge the digital, physical, and even biological worlds into a single entity referred to as the Cyber-Physical System (CPS) to enable innovative models of personalized production and servicing \cite{doi:10.1142/S2424862217500142}. In this regard, production machines should interact autonomously with each other with little (or no) human intervention to meet an ever-growing demand for efficiency, speed, quality, and sustainability. Therefore, the scientific and technological achievements that currently allow processing (e.g., sense, collect, store, and extract meaningful insights) massive amounts of data in an effective and cost-affordable way have become key enablers for Industry 4.0 \cite{doi:10.1080/09537287.2020.1810767}.

 Although the aim and scope of Industry 4.0 are well-known \cite{russmannIndustryFutureProductivity} in both industrial and scientific domains, several barriers have been identified that prevent its broad and smooth adoption \cite{RePEc:eee:proeco:v:224:y:2020:i:c:s092552731930372x}. These barriers range from skill deficits and mismatches \cite{brunelloSkillShortagesSkill2019} to industrial companies' lack of digital strategies, including poor competence in adopting new business models or a limited understanding of ethics and safety \cite{CHAUHAN2021124809}. While these issues are currently being addressed from a training and reskilling point of view \cite{Karacay2018}, it is fair to say that half of the picture is still hidden. Indeed, these intrinsic and extrinsic barriers only explain the situation from the industrial companies’ point of view but seem to neglect the technological and scientific perspectives \cite{doi:10.1080/09654313.2017.1327037}. Particularly, someone may wonder to what extent the developments and technical achievements that support Industry 4.0, which the academic/scientific domain has proposed in the fields \cite{russmannIndustryFutureProductivity}, have a practical, real-world application in an industrial context. Without a doubt, most of the published academic achievements have been tested under some laboratory environment aimed to mimic real-world conditions and passed peer-review processes and/or the scrutiny of a large community. However, the underlying question still holds: to what extent are these advancements solving real-world and real-operation Industry 4.0 challenges so companies can directly benefit from them?

Beyond the aforementioned intrinsic and extrinsic barriers \cite{CHAUHAN2021124809}, there is currently some paradox \cite{doi:10.1080/09654313.2017.1327037} since, despite the scientific progress in the field of Industry 4.0 enabling technologies, they fail to reach their primary target sector: the real-world industrial environments. Indeed, a common situation for many entrepreneurs nowadays is that they do not know which technologies are suitable and available for solving their challenges unless they already own the specific competencies (i.e., hard skills) for this challenge \cite{doi:10.1177/1465750320927359}. However, this does not happen predominantly, and even if they are aware of these technologies, they do not know how to apply them, a difficulty coming from higher education. In particular, academics cannot find real case studies to include in the educational process of digital technologies. Therefore, graduates and future engineers do not know how to utilize their knowledge and skills, thus, not being able to adapt to the ever-changing industrial environment.

It can be hypothesized that the overarching problem—beyond training and reskilling—shall be related to mutual understanding. That is a lack of a bidirectional communication environment where industry and academia could successfully interact and benefit from each other. On the one hand, industrial companies could describe their pains and challenges so that scientists could understand. In contrast, scientists from academia could use this environment to share, explain, and validate their achievements with industrial companies.

This paper aims to propose a knowledge exchange and communication environment for industry experts and academics/researchers, to enable effective communication between them when solving Industry 4.0 challenges (i.e., common understanding). Inspired by the Esperanto language, this communication environment aims to blend humanistic and engineering techniques applied to the technology, entrepreneurship, and industrial domains to facilitate the broad adoption of Industry 4.0. This proposal sought to be materialized into a structured and indexable collection of real-world industrial problems and their associated enabling technologies, called taxonomy. In this way, Industry 4.0 stakeholders from industry and academia could use this taxonomy as a meeting point to collaborate toward a smooth digital transformation of industries. 
This paper -as a continuation of the research work \cite{amadio2021building}- describes the rationale behind building this taxonomy which assorts more than 32 real-world industrial challenges and links them with 147 enabling technologies through their associated success stories. This effort was carried out within the framework of the Erasmus+ project PLANET4. The contributions of this paper are the following:
\begin{itemize}
\item State-of-the-art review on the main drivers of this communication gap between academia and industry in Industry 4.0.
\item Methodological approach to build a taxonomy of Industry 4.0 challenges and enabling technologies.
\item First version of the Industry 4.0 taxonomy and the corresponding web application.
\item Presentation of the taxonomy application to example industrial problems.
\end{itemize}

The remainder of this paper is organized as follows. Section \ref{state of the art} discusses the main causes that have led to this communication gap between academia and industry in the field of Industry 4.0 and reviews the most relevant attempts to mitigate it that have been proposed so far. Next, Section \ref{methodology} describes the methodological approach for building the proposed taxonomy. Section \ref{results and discussion} exhibits the obtained results, how this taxonomy was transformed into a usable web tool and its evaluation in two real scenarios. Finally, Section \ref{conclusions} concludes the paper, highlights the encountered work limitations, and proposes some future work directions.

\section{State of the art} \label{state of the art}

\subsection{The skill mismatch and gap of Industry 4.0} \label{the skill mismatch and gap of Industry 4.0}

Industry 4.0, as the primary bearer of technological advances in the industry nowadays, poses structural and organizational changes to firms. Industry 4.0-related technologies require firms to understand the possible opportunities and what they offer to tackle their business challenges. Although the Industry 4.0 implementation has mainly concerned large companies, the need for innovation activities in Small and Medium Enterprises (SMEs), which represent 99.8\% of enterprises in the EU, is increasing  \cite{executiveagencyforsmallandmediumsizedenterprises.AnnualReportEuropean2021}. This implementation process has already prompted modifications in business models adopted by SMEs, derived mainly from the companies' internal motivation and technological characteristics \cite{RePEc:eee:tefoso:v:132:y:2018:i:c:p:2-17}. However, whereas some SMEs can show a human and infrastructural predisposition to implement 4.0 technologies, many manifest limits in identifying new technologies due to the lack of specific experience in the human and managerial resources sectors \cite{KOWALKOWSKI201318}. 

Moreover, to successfully implement Industry 4.0, firms must adapt to the new requirements regarding skills. Whenever a structural change happens, old jobs and related required skills become obsolescent, leaving space for new jobs and their associated sets of new skills. In such a context, many workers and firms find themselves in a condition of skill mismatch \cite{restrepoStructuralUnemployment}.

The term “skill mismatch” defines situations where there are imbalances between the needed level or type of skills (or qualifications) in the labor market and the corresponding skills an individual possesses. It may be due to overeducation, undereducation, overskilling, underskilling, or skills shortage/ surplus/gap/obsolescence \cite{BriefingnoteSkillmismatchinEurope}. Too often, this mismatching occurs as “the result of poor adaptability of companies to long-term changes and transformation in their economies” \cite{RePEc:ssi:jouesi:v:8:y:2020:i:1:p:83-102}.

Digital skills related to Industry 4.0 concern the technical ability, comprehension, and use of AI, IoT and Data Science \cite{coskunAdaptingEngineeringEducation2019,pasko2022plan}. However, the I4.0 worker is now required to possess the organizational and personal capabilities for implementing those technological solutions in alignment with business needs (e.g., transversal skills, collaboration in different fields of expertise and interdisciplinary teams, often referred to as soft skills) \cite{hernandez-de-menendezCompetenciesIndustry2020,refId0}. Though the I4.0 implementation within companies concerns more technical aspects, they often overlook the socio-technical ones that put machines in a relationship with humans and the environment. It is foreseeable that a new probable professional with a broader understanding of the latest technologies that will appear in the 4.0 context will guide the company to adopt them \cite{DAVIES20171288}. Nevertheless, I4.0 technologies and needs are constantly evolving and thus necessary skills, making the problems of adoption and mismatch even more complicated. This process creates or widens the gap between the evolution of job profiles and their abilities and, consequently, how companies face it \cite{pinzoneJobsSkillsIndustry2017}. 

\subsection{The role of Academia in Industry 4.0} \label{the role of Academia in Industry 4.0}

In the implementation process of I4.0, universities can play a crucial role through training activities and technology/knowledge transfer. The former refers both to the higher education offer, which is preparing the future workforce, and the reskilling support that universities can offer to current company employees, acknowledging the difference in educational needs between the two (especially the balance between theory and practice in the learning process). This accounts for the so-called universities’ "first mission" - i.e., the qualification of human capital, i.e., education - \cite{fernandez_2019}, which makes academics responsible for the training of future generations of workers also in the I4.0 context \cite{10.1145/3452369.3463817}. The "second mission" refers to the production of new knowledge by the university, i.e., research duties. However, it is only in the complementary "third mission" that universities contribute to solving socio-economic problems by adopting an entrepreneurial mindset and a strategic plan that allows “the generation, use, application and exploitation of knowledge with external stakeholders and society in general” \cite{SECUNDO2017229}. 

University-industry collaboration tries to fulfill this third mission through knowledge and technology transfer and educational and research collaboration \cite{NSANZUMUHIRE2020120861}. For example, in the Triple Helix (academia-industry-government) model of innovation,  universities can provide support to SMEs through their technology transfer offices \cite{RICCI2021108234,kerryOpenInnovationTriple2016}. However, academia is usually perceived as a provider of teachings more than a technology provider, especially in SMEs with a weak willingness to cooperate with universities \cite{10.1093/reseval/rvx022}. Nevertheless, fewer companies rely on external cooperation with universities in providing a 4.0-ready education or development to train their employees \cite{stachovaExternalPartnershipsEmployee2019}.

From an industry perspective, collaborating with universities allows access and development of frontier technologies; from the university's perspective, the collaboration allows for expansion and leveraging of the resources to innovate. Moreover, in \cite{linBalancingIndustryCollaboration2017}, it has been shown how a moderate university-industry partnership has a positive impact as it leads to more (academic) innovation and facilitates knowledge transfer.

\subsection{Related works} \label{related works}

Possible solutions to tackle the skill mismatch in I4.0 are represented by training or reskilling the current internal workforce or hiring new workers that already have those skills \cite{behrendtLeveragingIndustrialIoT}. In each of these cases, knowing what is needed internally and what can help address it is crucial.

RAMI 4.0, SIMMI 4.0, Industry 4.0-MM, and M2DDM represent some of the frameworks and models available to assess the maturity level of I4.0 implementation \cite{koschnickRAMICombinesCrucial,weberM2DDMMaturityModel2017,7733413}. Two of the challenges highlighted in these models concern the companies’ training to develop adequate skills and the willingness to adopt 4.0 technologies \cite{hernandez-de-menendezCompetenciesIndustry2020}. However, differing from the mostly prescriptive nature of those frameworks, taxonomies (i.e., a classification methodology often used in Biology and Information Architecture) offer other - primarily descriptive - systematization models. They are elaborated to make sense of the landscape of technologies, business models or strategies involved in I4.0. 

\definecolor{LightCyan}{rgb}{0.88,1,1}
\definecolor{LightGray}{rgb}{0.94,0.94,0.94}


\begin{table*}
\centering
\caption{Comparison of related works’ taxonomies}
\resizebox{\textwidth}{!}{%
\begin{tabular}{clcclc} 
\hline
Paper                  & Key Topics                                                      & Technology-focused & Industry-focused & Methodology                                                                                                                                           & Taxonomy presentation  \\
\hline 

\rowcolor{LightGray}
Lagorio et al. \cite{lagorioTaxonomyTechnologiesHumanCentred2021}    & \hspace*{-2mm}\begin{tabular}{l}Logistics 4.0\\Human-centered logistics\end{tabular} &\cmark                                                                                  &\cmark                                                                                & \hspace*{-2mm}\begin{tabular}{l}Deductive approach\\Literature review\end{tabular}                                              & Table                                                            \\
Weking et al. \cite{WEKING2020107588}     & Industry 4.0 BMs                                                                                                                & \xmark                                                                                  &\cmark                                                                                & \hspace*{-2mm}\begin{tabular}{l}Inductive and~deductive approach\\Case studies analysis\\Taxonomy evaluation\end{tabular}        & Table                                                           \\
\rowcolor{LightGray}
Wagire et al. \cite{wagireAnalysisSynthesisIndustry2019}     & Industry 4.0 research dynamics                                                                                                  &\cmark                                                                                  &\cmark                                                                                & Latent Semantic Analysis                                                                                                                                                   & Pie / Table                                                      \\
Cañas et al. \cite{canasImplementingIndustryPrinciples2021}      & Industry 4.0 implementation                                                                                                     &\cmark                                                                                  &\cmark                                                                                & Literature review classification                                                                                                                                           & Table                                                            \\
\rowcolor{LightGray}
de Moura et al. \cite{demouraTechnologiesIndustrialInternet2021}   & IIoT and Edge Computing                                                                                                         &\cmark                                                                                  & \xmark                                                                                                                                                                 & Not provided                                                                                                                                                               & Textual                                                          \\
Nazarov \& Klarin \cite{nazarovTaxonomyIndustryResearch2020} & Industry 4.0 research                                                                                                           &\cmark                                                                                  &\cmark                                                                                & \begin{tabular}[c]{@{}l@{}}Thematic analysis\\Scientometric analysis\end{tabular}                                                                                          & Table / Cluster
visualization                                     \\
\rowcolor{LightGray}
Oztemel \& Gursev \cite{oztemelTaxonomyIndustryRelated2020} & Industry 4.0 overall landscape                                                                                                  &\cmark                                                                                  &\cmark                                                                                & Literature review                                                                                                                                                          & Hierarchical tree                                                \\
Latino et al. \cite{9523790}     & Agriculture 4.0                                                                                                                 &\cmark                                                                                  &\cmark                                                                                & \hspace*{-2mm}\begin{tabular}{l}Frequent-terms analysis\\Experts focus group\\Inductive approach\end{tabular}                   & Hierarchical tree                                                \\
\rowcolor{LightGray}
da Silva et al. \cite{teschdasilvaLookingEnergyLens2020}   & \hspace*{-2mm}\begin{tabular}{l}Smart manufacturing\\Energy consumption\end{tabular} &\cmark                                                                                  &\cmark                                                                                & Systematic literature review                                                                                                                                                & Hierarchical tree                                                \\
Zhou et al. \cite{zhouSurveyVisualizationSmart2019}       & \hspace*{-2mm}\begin{tabular}{l}Smart manufacturing\\Visualization\end{tabular}      &\cmark                                                                                  &\cmark                                                                                & Literature review                                                                                                                                                          & Table                                                            \\
\rowcolor{LightGray}
Raptis et al. \cite{8764545}     & Industry 4.0 Data Management                                                                                                    &\cmark                                                                                  &\xmark                                                                                & Literature review                                                                                                                                                          & Hierarchical tree                                                \\
Zonta et al. \cite{zontaPredictiveMaintenanceIndustry2020}      & Industry 4.0 Predictive
maintenance                                                                                             &\cmark                                                                                  &\cmark                                                                                & \hspace*{-2mm}\begin{tabular}{l}Systematic literature review\\NLP clustering\\Mapping of clusters\end{tabular}                   & Hierarchical tree                                                \\
\rowcolor{LightGray}
Berger et al. \cite{bergerAttacksIndustrialInternet2020}     & Cyber-attacks in IIoT                                                                                                           &\cmark                                                                                  &\xmark                                                                                                                                                                &  \hspace*{-2mm}\begin{tabular}{l}Systematic literature review~\\Inductive and deductive approach\\Experts interviews\end{tabular} & Table                   \\ \hline                                      
\end{tabular}
}

\label{tab:1}
\end{table*}


In Table \ref{tab:1}, taxonomies elaborated in related works are compared as follows:
\begin{itemize}
\item the "Key topics" column refers to the main topics addressed by each taxonomy;
\item  the "Technology-focused" column verifies if the taxonomy covers technical aspects of I4.0;
\item the "Industry-focused" column refers to the application or implementation of I4.0 aspects at the business, managerial, or strategic level at large; 
\item  the "Methodology" column lists the procedures applied in the development of the taxonomy; 
\item the "Taxonomy presentation" column refers to the final appearance of the taxonomy.
\end{itemize}

Specific aspects of I4.0 have been covered in taxonomies. For example, IIoT and EC are addressed by taxonomy in the study of de Moura et al. \cite{demouraTechnologiesIndustrialInternet2021} and smart manufacturing applications for visualization technologies in Zhou et al. \cite{zhouSurveyVisualizationSmart2019}. 

The study by Lagorio et al. \cite{lagorioTaxonomyTechnologiesHumanCentred2021} presents a taxonomy to denote the relationships between technologies and human factors in  Logistics 4.0 through a literature review and a deductive approach to identify its 3 main categories and 10 dimensions. 

Da Silva et al. \cite{teschdasilvaLookingEnergyLens2020} examined energy consumption in the smart industry. Through a systematic literature review of 53 articles, the authors propose a taxonomy in a hierarchical tree that considers "Goals", "Concerns", and "Deployment", i.e., I4.0 technologies. Thus, the resulting taxonomy covers the challenges, needs, and technologies related only to the topic of energy in I4.0.

Raptis et al. \cite{8764545} analyzed the state-of-the-art research on data management in I4.0 through a literature review. Then, they provide a comprehensive and holistic taxonomy - presented in a hierarchical tree - based on 316 articles reviewed and considering all the main enablers, such as "data enabling technologies" and "data-centric services". 

An overview of predictive maintenance in the context of I4.0 is given by the study of Zonta et al. \cite{zontaPredictiveMaintenanceIndustry2020}. After a systematic literature review on the subject, the authors propose a taxonomy built through applying natural language analysis of the articles, clustering the more frequent terms, and mapping them in a hierarchical order.

Weking et al. \cite{WEKING2020107588} presents a taxonomy of Industry 4.0 business models for innovation deriving from the analysis of 32 industrial case studies. Their study proposes a well-defined methodology - composed of an iterative process - resulting from applying taxonomies building guidelines given by Nickerson et al. \cite{nickersonMethodTaxonomyDevelopment2013}.

Likewise, Berger et al. \cite{bergerAttacksIndustrialInternet2020} built a taxonomy on cyber-attacks in IIoT after a literature review of the relevant works and following the taxonomy-building methodology elaborated by Nickerson et al. \cite{nickersonMethodTaxonomyDevelopment2013}. In such a manner, the authors follow a 4-phases iterative process to develop the taxonomy, which involves a systematic literature review and validation with industry experts.

Through a Latent Semantic Analysis approach on 503 research papers on I4.0, Wagire et al. \cite{wagireAnalysisSynthesisIndustry2019} identify the main research areas and themes addressed by academia. In particular, the resulting taxonomy is represented in a 3 levels pie graphic. Each of them denotes a category (e.g., "Principal research areas", "Minor research themes", and "Major research themes"). Despite the immediacy of this type of representation, understanding its scheme could be misleading as the order of the circles may suggest a tree-like structure (and, therefore, a subsumptive classification scheme).  

Similarly, Nazarov and Klarin \cite{nazarovTaxonomyIndustryResearch2020} provide a taxonomy of the literature on I4.0 research utilizing the scientometric analysis to help identify “top trending and top articles”. Thus, the taxonomy provided results from identifying clusters of the domains addressed in the research on I4.0, their top terms, and their top articles. Although the taxonomy provides insights from both academia and the industry, aiming to propose a holist system view on I4.0, the results appear to be aimed more at an audience of researchers than at decision-makers and other stakeholders. 

Cañas et al. \cite{canasImplementingIndustryPrinciples2021} tried to clarify the intrinsic confusion that I4.0 (its technologies, conceptual frameworks, etc.) bears, proposing a taxonomy of the design principles for its implementation after the review of 130 papers. Notwithstanding the adequate amount of publications examined, the authors seem unable to translate the results of the classification process into a proper taxonomy presentation, presenting them only in a summary table with minimal categorization efforts. 

Oztemel and Gursev \cite{oztemelTaxonomyIndustryRelated2020} performed a literature review on 620 papers to elaborate an overall taxonomy of I4.0 covering 4 main aspects - namely "Strategic view", "Managerial view", "Technical view", and "Human Resource view" -. For each of these 4 main aspects, the authors identified other sub-categories, classifying them in hierarchical tree order. In this way, the authors succeed in analyzing both the business and the technological side of I4.0. However, it is not clear how these aspects - and their sub-categories - are mapped between each other.

Latino et al. have investigated state of the art on I4.0 implementation in agriculture (Agriculture 4.0) \cite{9523790}. The authors propose a taxonomy of the technologies, processes, issues, and aims involved in Agriculture 4.0 by reviewing and analyzing 1338 studies. The taxonomy is the result of comparing the more frequent terms identified in those studies and the thematic clusterization by an inductive approach done by a focus group of experts. Finally, the taxonomy is presented in a hierarchical tree.

Most of the examined taxonomies consider both business and technologies. However, whereas the technologies are often described in more detail, there is a general lack of specificity in identifying business-related categories or vice versa. Except for some taxonomies addressing particular aspects of I4.0, the vast majority suffer from this lack of specificity in identifying categories. Furthermore, the connection between these two sides is not always clear, leaving a gap in identifying the right technology for a specific issue.

In this study, far from refuting all previous works, we propose a bird-eye view of the I4.0 landscape, trying to keep its needs, theories, and practices together. This holistic perspective could provide valuable and operational information to all stakeholders involved in decision-making practices, bridging the gap between real-world and specific industry needs and the technological solutions that can tackle them to ease the progression towards implementing 4.0 principles. This led us to the following research questions:

\begin{itemize}
 \setlength{\parskip}{0pt}
  \setlength{\itemsep}{0pt plus 1pt}
   \item[] \textbf{RQ1:} \textit{“What are the main needs and challenges that industries/businesses face nowadays?”}
\item[] \textbf{RQ2:}  \textit{“What are the enabling technologies behind the solutions capable of solving industrial challenges?”}
\item[] \textbf{RQ3:} \textit{“How can we link those business challenges and enabling technologies to create a tool of shared knowledge to better implement Industry 4.0?”}
\end{itemize}

To address these research questions, we propose the building of two taxonomies linked together.

\section{Methodology} \label{methodology}

\subsection{Multivocal Literature Review Process} \label{multivocal literature review process}

\begin{figure}
\centering
\includegraphics[width=\columnwidth]{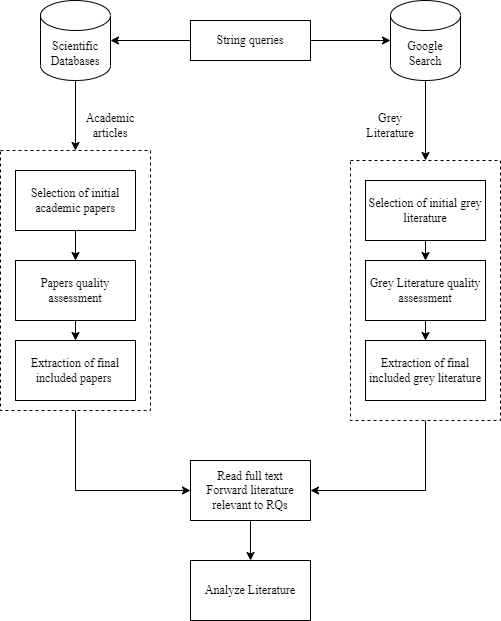}
\caption{Search process overview}
\label{fig:1}
\end{figure}

To answer thoroughly and reliably the defined research questions, the internal firm perspective had to be considered besides the research community's publications since many of them range in theory with no applicability in practice \cite{rosing}. In other words, we had to analyze scientific research papers and writings edited by people or organizations directly involved in the examined topic, called grey literature (GL). For this reason, a Multivocal Literature Review (MLR) was conducted to analyze both state-of-the-art and state-of-practice in the industrial sector \cite{ogawaRigorReviewsMultivocal1991}. The proposed methodology is illustrated in Fig. \ref{fig:1}.

Regarding academic literature research, to collect scientific publications, we employed the most well-known scientific search engines available online: Google Scholar, Scopus, ResearchGate, IEEE Xplore, and ScienceDirect.  The search engines were queried, making use of an API interface which permits the utilization of keywords, filters, and logic conditions to select the appropriate articles. The string queries have to be defined in such a way as to avoid rejecting articles that refer to challenges or technologies. For this reason, synonyms of digital transformation, change management, and knowledge innovation for the business part were used, while for the technologies side, some key I4.0 technologies, like wireless sensor networks (WSNs), IoT, CPSs, and smart factories, were listed. Each technology-related term was queried using the logic operator “OR” due to the high possibility of these technologies’ presence in the articles compared to more uncertain business categories. To this end, a data extraction pipeline was created. Different terms from the technology and business dictionary were combined to compose the query and extract the results from the scientific databases. A list of query samples can be viewed in Table \ref{tab:2}.


\begin{table*}
\centering
\caption{String queries used in academic and mainstream search engines}
\begin{tabular}{c|p{0.52\linewidth}cl|c}
\hline
Search Engine & \multicolumn{3}{c|}{Query} & Results \\ \hline
 \multirow{8}{*}{\rotatebox[origin=c]{0}{\begin{tabular}{p{1.5cm}}\centering Academic Literature: Scopus \end{tabular}}} &  \multirow{8}{*}{\begin{tabular}{p{0.94\linewidth}}(\centering “{\em Internet of Things}” {\bf OR} “{\em IoT}” {\bf OR} “{\em Smart Factory}” {\bf OR} “{\em Industry 4.0}” {\bf OR} “{\em Cyber Physical System}” {\bf OR} “{\em Wireless Sensor Network}”)\end{tabular}} &   \multirow{8}{*}{\centering \hspace*{-8mm}{\bf AND}}  & \multirow{2}{*}{ {\em Change Management}} & \multirow{2}{*}{\centering  780}\\
 &  &  & & \\
  &  &  & \cellcolor{LightGray} &\cellcolor{LightGray}  \\
 &  &  & \cellcolor{LightGray} \multirow{-2}{*}{ {\em Digital Management}} & \cellcolor{LightGray} \multirow{-2}{*}{\centering 762}\\
 &  &  & \multirow{2}{*}{ {\em Innovation Management}} & \multirow{2}{*}{\centering 739}\\
 &  &  & & \\
 &  &  & \cellcolor{LightGray} &\cellcolor{LightGray}  \\
  &  &  & \multirow{-2}{*}{\cellcolor{LightGray} {\em Technological Management}} &  \multirow{-2}{*}{\cellcolor{LightGray} \centering 542}\\
  \hline
  \hline 
   \multirow{4}{*}{\rotatebox[origin=c]{0}{\begin{tabular}{p{1.5cm}}\centering Grey literature: Google \end{tabular}}} & \multicolumn{3}{c|}{\multirow{2}{*}{“{\em Industrial}” {\bf AND} (“{\em Problems}” {\bf OR} “{\em Needs}” {\bf OR} “{\em Issues}” {\bf OR} “{\em Challenges}”)}} & \multirow{2}{*}{107}\\
    &  &  & & \\
    & \cellcolor{LightGray} & \cellcolor{LightGray}  &\cellcolor{LightGray} & \cellcolor{LightGray} \\
   & \multicolumn{3}{c|}{\multirow{-2}{*}{\cellcolor{LightGray} \begin{tabular}{p{0.75\linewidth}}(\centering “{\em Industry 4.0}” {\bf OR} “{\em Smart Manufacturing}” {\bf OR} “{\em Smart Industry}”) {\bf AND} (“{\em Solutions}” {\bf OR} “{\em Technologies}” {\bf OR} “{\em Examples}” {\bf OR} “{\em Usecases}” {\bf OR} “{\em Projects}” {\bf OR} “{\em Products}” {\bf OR} “{\em Case studies}” {\bf OR} “{\em News}”)\end{tabular}}} &  \multirow{-2}{*}{\cellcolor{LightGray}126} \\

  \hline
  
\end{tabular}
\label{tab:2}
\end{table*}

The total number of articles exported from this search was 2823. However, before using studies in the review, they were appraised critically for quality and risk of bias. Consequently, the PRISMA (Preferred Reporting Items around Systematic Reviews and Meta-Analyzes) framework was utilized \cite{pagePRISMA2020Explanation2021}. It is a methodology based on formulated inclusion and exclusion criteria that systematically assesses the quality of chosen papers and either includes or excludes them. The flow chart of PRISMA is based on four stages:

\begin{itemize}
\setlength{\parskip}{0pt}
  \setlength{\itemsep}{0pt plus 1pt}
   \item \textbf{Identification:} Identify the papers based on search strategies.
\item \textbf{Screening:} Use the inclusion or exclusion criteria and the quality checklist to exclude irrelevant papers.
\item \textbf{Eligibility:} Prioritize using the quality checklist to find the papers' eligibility.
\item \textbf{Included:} Review the papers critically to address the aim of the current study.
\end{itemize}

During this process (Fig. \ref{fig:new}), duplicate articles were initially removed from the dataset, as well as non-English writings. Subsequently, the eligibility of the selected papers was checked based on the quality checklist items and the exclusion criteria, which are: a) date of publication, b) availability of the abstract, c) access to the full text, d) relevancy with the predefined scope, e) possibility of applying the proposed technological solutions to real industrial problems and f) comprehensive analysis of the results. Finally, the exported number of articles included in this study was 365.



\tikzset{
    mynode/.style={
        draw, rectangle, align=center, text width=5cm, font=\small, inner sep=3ex},
    mylabel/.style={
        draw, rectangle, align=center, rounded corners, font=\small\bf, inner sep=2ex, 
        fill=cyan!30, minimum height=3.8cm},
    arrow/.style={
        very thick,->,>=stealth}
}

\begin{figure}
\centering

\resizebox{0.5\textwidth}{!}{%
\begin{tikzpicture}[
    node distance=9mm and 10mm,
    start chain=going below,
 mynode_left/.style = {
        draw, rectangle, align=center, text width=2cm, 
        font=\small, inner sep=1ex, outer sep=0pt,
        on chain},
 mynode/.style = {
        draw, rectangle, align=center, text width=3cm, 
        font=\small, inner sep=3ex, outer sep=0pt,
        on chain},
 mynode_wide/.style = {
        draw, rectangle, align=center, text width=7cm,
        font=\small, inner sep=1ex, outer sep=0pt,
        on chain},
mylabel/.style = {
        draw, rectangle, align=center, rounded corners, 
        font=\small\bfseries, inner sep=1ex, outer sep=0pt,
        fill=cyan!30, minimum height=38mm,
        on chain},
every join/.style = arrow,
     arrow/.style = {very thick,-stealth}
                    ]  
\node (n1a) [mynode_wide, xshift=15cm]    {Records identified through database searching\\ (n = \textbf{2823})};
\node (n2)  [mynode_wide, below=of n1a]   {Records after duplicates removed\\ (n = \textbf{2405})};
\node (n3)  [mynode_left, xshift=-2.5cm]   {Records screened\\ (n = \textbf{2405})};
\coordinate (n2_shifted) at ([xshift=-2.5cm]n2.south);
\draw[arrow] (n2_shifted) -- (n3); 
\node (n4)  [mynode_left,join]   {Full-text articles accessed 
                                            for eligibility\\ (n = \textbf{1536})};
\node (n5)  [mynode_left,join]   {Studies included in qualitative synthesis\\ (n = \textbf{365})};
\node (n6)  [mynode_left,join]   {Studies included in quantitative synthesis\\ (n = \textbf{365})};
\node (n3r) [mynode,right=of n3]    {{Records excluded based on publication date and writing language\\ (n = \textbf{869})};};
\node (n4r) [mynode,right=of n4, yshift=-2.9cm]    {\vspace*{-0.5cm}

 \raggedright Full-text articles ex- cluded, with reasons: 
            \begin{itemize}
                \item Availability of the abstract (n = \textbf{25})
                \item Access to the full text (n = \textbf{8})
                \item Relevancy with the predefined scope (n = \textbf{395})
                \item Possibility of applying the proposed technological solutions to real industrial challenges (n = \textbf{580})
                \item Comprehensive analysis of the results (n = \textbf{163})
            \end{itemize}
            \vspace*{-0.6cm}\textcolor{white}{.}};
\draw[arrow] (n1a) -- (n2);
\draw[arrow] (n3) -- (n3r);
\coordinate (n4rshifted) at ([yshift=2.9cm]n4r.west);
\draw[arrow] (n4) -- (n4rshifted);  
    \begin{scope}[node distance=7mm]
\node[mylabel,below left=0mm and 11mm of n1a.north west, minimum height=2.8cm]
                {\rotatebox{90}{Identification}};
\node[mylabel, minimum height=1.8cm, yshift=0cm]  {\rotatebox{90}{Screening}};
\node[mylabel, minimum height=2.2cm, yshift=0.2cm]  {\rotatebox{90}{Eligibility}};
\node[mylabel, minimum height=5.0cm, yshift=0.1cm ]  {\rotatebox{90}{Included}};
    \end{scope}
\end{tikzpicture}
}

\caption{PRISMA Flow Diagram}
\label{fig:new}
\end{figure}
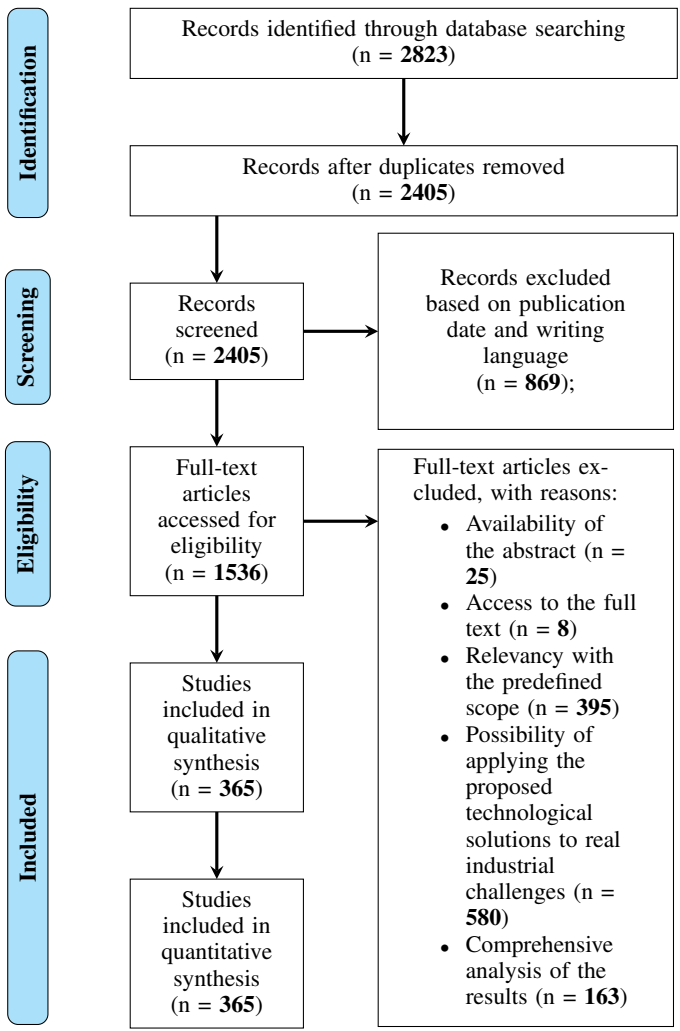


For collecting and evaluating the GL, we followed the guidelines provided by Garousi \cite{fabf882fbc0f4e3cabb05b86bafddc8c}. At first, a general web search engine (i.e., Google) instead of academic databases was employed. At the same time, a generic query string was used to extract unbiased results, containing only the main keywords of the subject (Table \ref{tab:2}). 

When theoretical saturation was reached, i.e. when no new concepts emerged from the search results, the GL search was terminated. After collecting the initial number of sources, it was necessary to evaluate their quality as they are usually not subject to review before publication. On account of this, the "ACCODS" checklist proposed by Tyndall was used, which is an evaluation tool of the quality of grey information according to the following eligibility criteria \cite{tyndallHowLowCan2008}:

\begin{itemize}
\setlength{\parskip}{0pt}
  \setlength{\itemsep}{0pt plus 1pt}
   \item \textbf{Authority:} Who is responsible for the intellectual content?
\item \textbf{Accuracy:} Is it representative of work in the field?
\item \textbf{Coverage:} Are any limits clearly stated?
\item \textbf{Objectivity:} Is there any bias?
\item \textbf{Date:} Have key contemporary material been included?
\item \textbf{Significance:} Is the source meaningful?
\end{itemize}

The final number of GL studied in this work was carried out, with 76 different sources chosen for the industrial challenges and the enabling technologies.

\subsection{Theoretical framework of taxonomies and their interoperability} \label{theoretical framework of taxonomies and their interoperability}

Once the literature to be studied has been determined through the MLR process, described in section \ref{multivocal literature review process}, the problems faced by the manufacturing industry and their technological solutions were extracted by reading each source. Subsequently, the identified business problems and technologies were organized in a taxonomic hierarchy created from intuitive and generic logic and in-depth analysis of articles, surveys, business reviews, technical reports, case studies, and whitepapers with the same objective \cite{bajicIndustryImplementationChallenges2021,vuksanovichercegChallengesDrivingForces2020}.

The term taxonomy can be described as a scheme of classification of concepts to represent the relationships that they have between them. Usually, a taxonomy defines only a narrow set of relationships (parent-child or hierarchical). However, in  Information Architecture, the term is often used to describe a general form for "organizing concepts of knowledge" \cite{Hedden_2016}. Indeed, a broader definition of taxonomy defines it as the logical structure that gives meaning to what is being presented \cite{covert2014make} or "as a knowledge organization system or knowledge organization structure" \cite{Hedden_2016}. Here "taxonomy" will be used in the narrower sense of the term, distinguishing it from thesauruses and ontologies (in which there are more relationship types).

Each taxonomy element is defined by its concept, i.e., the idea or the thing  identified, and one or more terms (synonyms), i.e., the label that describes the concept. Usually, only one preferred term will designate a concept. As already mentioned, a taxonomy presents a hierarchical structure that categorizes elements of the same domain in a subsumptive manner, i.e., a concept of a higher-order level will be broader and more generic, and a concept of a lower level will be narrower and more specific \cite{NISO_2010}.

As previously anticipated, other classification schemes have more typologies of the relationship between their concepts, thesauruses and ontologies. Unlike taxonomies, thesauruses are controlled vocabularies "arranged in a known order and structured so that the various relationships among terms are displayed clearly and identified by standardized relationship indicators" \cite{NISO_2010}, i.e., providing three types of standard relationships (hierarchical, associative, and equivalence).  Finally, ontologies provide non-standard, domain-specific relationships defined by the ontology creators \cite{Hedden_2016}.

For the purpose and scope of this research, we preferred to use the simpler version of knowledge organization structure: the taxonomy. This choice was led by the goal of classifying the main concepts concerning the challenges and the technologies of I4.0 in a tree structure by employing only the most used terms that usually identify these concepts in academia and the business world.

Given the twofold scope of this taxonomy, two separate structures would describe the domains of the challenges and technologies related to I4.0. It was, therefore, necessary to investigate how these two structures, which precisely identify two different taxonomies, could become interoperable. Hedden explains that combining two or more taxonomies is feasible by utilizing 3 possible procedures  \cite{Hedden_2016}: Integration, Merging and Mapping.

Merging refers to the action of combining two or more taxonomies into a single one focusing on the equivalence relationship among their concepts (thus making the original structures disappear). Mapping refers to linking taxonomies’ concepts with each other (i.e., establishing a semantic correspondence between them) and maintaining the original structures. Integration permits the combination of more taxonomies into a new master taxonomy. As the two taxonomies about I4.0 business challenges and enabling technologies had to be linked but remain distinct in their structures, we first excluded the merging procedure. On the other hand, integrating the taxonomies would have led to encapsulating one taxonomy into the other, for example, adding it as a new branch. Finally, mapping was either not possible because there is no exact match between each other concepts (e.g., Smart PPE is not a synonym for IoT and vice versa).

Nevertheless, mapping is only a specific type of link, which, in its broadest sense, appears to be just the type of link that insists between the concepts of the two taxonomies in general. In this work, the studied articles and documents are considered the content that establishes the connection between two or more concepts belonging to the two taxonomies. This (associative) relationship would allow a bi-directional linking between their concepts.

\subsection{Methodology for the taxonomy development}

Following the existing design standards for taxonomy building \cite{NISO_2010} as guidelines for this work, we also incorporated some of the methodologies coming from the field of Information Systems, like the one presented by Nickerson et al. \cite{nickerson_method_2013} and the updated version by Kundisch et al. \cite{kundisch_update_2021}. More precisely, we have considered the principles from the Information Architecture field as the general philosophy and structure for the taxonomy building and presenting. In contrast, the area of Information Systems has inspired us in the taxonomy development process. This allowed us to determine a precise methodology to develop the taxonomy, building it as a controlled vocabulary of terms.

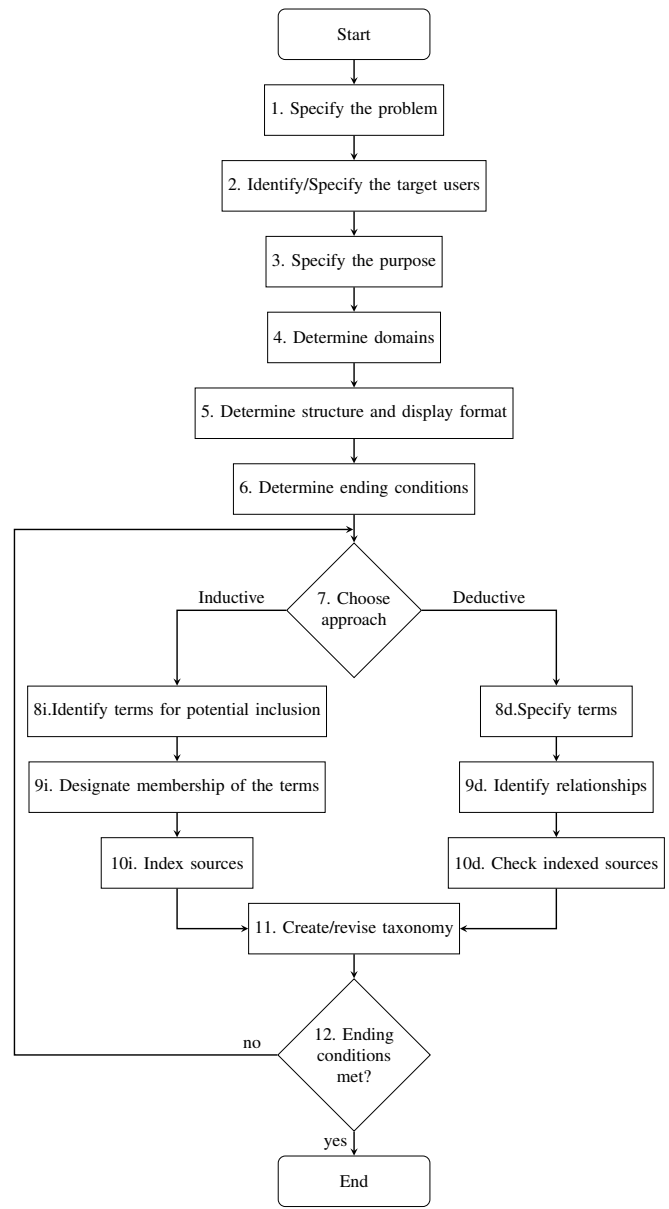
\begin{figure}
\adjustbox{max width=0.5\textwidth}{%
\centering
\tikzstyle{startstop} = [rectangle, rounded corners, minimum width=3cm, minimum height=1cm,text centered, draw=black]
\tikzstyle{decision} = [diamond, draw, text width=5.5em, text badly centered, inner sep=0pt]
\tikzstyle{process} = [rectangle, minimum width=3cm, minimum height=1cm, text centered, draw=black]
\tikzstyle{arrow} = [thick,->,>=stealth]

\begin{tikzpicture}[node distance = 2cm]
\node (start) [startstop] {Start};
\node (step1) [process, below of=start, yshift=.5cm] {1. Specify the problem};
\node (step2) [process, below of=step1, yshift=.5cm] {2. Identify/Specify the target users};
 
\node (step3) [process, below of=step2, yshift=.5cm] {3. Specify
the purpose};
\node (step4) [process, below of=step3, yshift=.5cm] {4. Determine domains};
\node (step5) [process, below of=step4, yshift=.5cm] {5. Determine structure and display format};
\node (step6) [process, below of=step5, yshift=.5cm] {6. Determine ending conditions};
\node (dec1) [decision, below of = step6, yshift=-0.4cm] {7. Choose approach};

\node (step8i) [process, left of=dec1, xshift=-1.5cm, yshift=-2cm] {8i.Identify terms for potential inclusion};
\node (step8d) [process, right of=dec1, xshift=2cm, yshift=-2cm, align=center] {8d.Specify terms};

\node (step9i) [process, below of=step8i, yshift=.5cm] {9i. Designate membership of the terms};
\node (step9d) [process, below of=step8d, yshift=.5cm] {9d. Identify relationships};

\node (step10i) [process, below of=step9i, yshift=.5cm] {10i. Index sources};
\node (step10d) [process, below of=step9d, yshift=.5cm] {10d. Check indexed sources};
 
\node (step11) [process, below of=step6, yshift=-6.7cm] {11. Create/revise taxonomy};
\node (dec2) [decision, below of=step11, yshift=-0.5cm] {12. Ending conditions met?};

\node (end) [startstop, below of=dec2, yshift=-0.5cm] {End};
 
\draw [arrow] (start) -- (step1);
\draw [arrow] (step1) -- (step2);
\draw [arrow] (step2) -- (step3);
\draw [arrow] (step3) -- (step4);
\draw [arrow] (step4) -- (step5);
\draw [arrow] (step5) -- (step6);
\draw [arrow] (step6) -- (dec1);

\draw [arrow] (dec1) -| node[near start, above] {Inductive} (step8i);
\draw [arrow] (dec1) -| node[near start, above] {Deductive} (step8d);

\draw [arrow](step8i) -- (step9i);
\draw [arrow](step8d) -- (step9d);

\draw [arrow](step9i) -- (step10i);
\draw [arrow](step9d) -- (step10d);

\draw [arrow](step10i) |- (step11);
\draw [arrow](step10d) |- (step11);

\draw [arrow] (step11) -- (dec2);

\draw [arrow] (dec2) -- node[anchor = east] {yes} (end);
\draw [arrow]  (dec2.west) node[anchor = south, xshift=-.5cm] {no}-- +(-5.2,0) |- +(1.5,10.4);
\end{tikzpicture}
}
\caption{Taxonomy development methodology}
\label{fig:3}
\end{figure}

Fig. \ref{fig:3} illustrates the process adopted in the development of the taxonomy. The first three steps help in identifying the "why" ("Specify the problem"), what is the intention ("Specify the purpose"), and for whom this new taxonomy is intended ("Identify/Specify the target users"). These steps were already addressed in the introduction and state-of-the-art. 

The following three steps (4-5-6) help better understand and identify how the taxonomy is intended to be developed. In our case of a twofold taxonomy, the domains determined are two instead of only one. This poses the problem of what is the object we developed: one or two taxonomies? As already argued previously, from a purely theoretical perspective having two separated domains with their respective structures requires us to consider them as two separated taxonomies. Nevertheless, the purpose of viewing the two domains together and the process brought us to develop the taxonomies jointly and consider them as a single integrated taxonomy for I4.0 solutions. The two domains are industrial needs and the I4.0 enabling technologies (i.e., the technologies that constitute the foundations of I4.0 solutions). 

As for the structure and display format, given the intention to build a hierarchy and made explicit primarily by using the term taxonomy in a narrow sense, the choice fell on a tree structure. The ending conditions provide a means of understanding when the taxonomy construction process can be declared finished or reiterated. Like the rest of the methodology, the adopted ending requirements were inspired by the criteria indicated by Nickerson et al. \cite{nickerson_method_2013} and the ANSI/NISO standards \cite{NISO_2010} both. The ending criteria determined are:

\begin{itemize}
    \setlength{\parskip}{0pt}
    \setlength{\itemsep}{0pt plus 1pt}
    \item The terms must be consistent with the domain and the structure (e.g., ensure that there are no duplicates, wrong hierarchical relationships, etc.). 
    \item The terms must be clear and validated according to reference texts, technical dictionaries, expert advice or common usage.
    \item No term must be without source/s or children.
    \item All the sources identified must be associated (indexed) with one or more terms.
    \item The terms must have definitions.
    \item No new terms or sources were added in the last iteration (a new term requires old sources to be checked for possible matchings, a new source could identify new terms).
\end{itemize}

The following steps (from 7 to 12) are intended as the iterative part of the methodology. The work by Nickerson et al. and the ANSI/NISO standards refer to 2 possible courses of action: the deductive (or conceptual-to-empirical) approach and the inductive (or empirical-to-conceptual) approach. Although the two methodologies differ in some of their details -i.e., in Nickerson, the deductive process does not contemplate the examination of the objects (the literature in our case), while in the ANSI/NISO standards is the first step- they could be integrated as needed. For example, in the first iteration of the taxonomy, we followed a deductive approach as indicated by ANSI/NISO standards, beginning with the extraction of terms with the assistance of topic modelling algorithms. Afterwards, we first classified those terms, starting from the broader concepts to narrower (top-down approach), based on their use in the sources, common usage in their fields, and expert advice. 

The inductive approach involves adding new relevant terms encountered in the examined literature, first identifying the narrower concepts and then finding the broader ones (bottom-up approach). As stated in the ending conditions, each source analyzed must be associated or, more appropriately, indexed under at least a term to achieve the goal of a taxonomy based on real-world, data-powered, and proven applications. As the inductive approach is based only on the analysis of objects (the literature), indexing the sources was mainly included in this procedure branch. During this process, the following issue appeared: regarding the GL, due to its heterogeneous nature, many sources deal with many topics (i.e., solving various industrial challenges with disparate technologies), presenting them in the same writing. Therefore, the above sources have been indexed multiple times (with different reference numbers), each corresponding to one challenge with the related technologies used to solve it. However, when iterating on terms with the deductive approach was also necessary to review references already indexed. After each iteration, the ending conditions were tested to verify the development status. A new iteration is required if it does not meet the criteria; otherwise, the process is considered ended.

\section{Results and Discussion} \label{results and discussion}

\subsection{Overview of the industrial challenges} \label{overview of the industrial challenges}

This section provides an overview of the challenges faced by the industry and the technologies that can help address these challenges. 

Answering the research question \textbf{RQ1}: "\textit{What are the main needs and challenges that industries/businesses face nowadays?}" we proposed to group the industrial challenges.

According to Kinkel et al., the innovation activities in
the manufacturing industry are distinguished between process
and product innovation, comprising both
technological and organizational innovations \cite{RePEc:zbw:fisibu:33e}. Therefore, industrial needs are divided into two main categories, \textbf{Process Optimization} and \textbf{Product Innovation}. The identified subcategories of the former category are:
\begin{itemize}
    \item \textit{Equipment and Process Efficiency Improvement}: Concerns activities related to the manufacturing ecosystem connectivity for alerting, running data analytics processes, and all the maintenance processes that ensure continuous readiness and operation of the industrial equipment without unplanned disruptions. It also involves initiatives for a) production planning and scheduling, b) employing industrial units that work independently without human intervention, c) legacy systems modernization and retrofitting, and d) commissioning processes.
    
    \item \textit{Worker security improvement and accident prevention}: Problems in this area are related to the use of effective means for workers' personal and collective protection.

    \item \textit{Supply Chain}: Needs that ensure a smooth flow of information and materials inside the company between individual production departments (internal suppliers and customers) and between the company and its external suppliers and customers.

   \item \textit{Mass Customization}: Individualization of the production process while remaining profitable for the enterprise. That is to implement technologies allowing for unlimited adaptation of products to individual customer requirements without affecting the unit production costs.

   \item \textit{Quality Assurance}: It is connected with the necessity to monitor the operation of machines and process parameters to identify quality disturbances and take immediate corrective and preventive actions.

    \item \textit{Sustainable industrial environment}: Is related to the appropriate management of industrial processes purposing of reducing energy consumption and minimizing/utilizing wastes, among others, from an environmental, economic and social perspective.
    
    \item \textit{Employee training and Assistance in worker’s tasks}: The use of appropriate technologies that help prepare employees to perform a new job and support them in implementing current jobs (e.g., product design, manual assembly, inspection activities, maintenance tasks, and order picking), minimizing the risk of human errors.

    \item \textit{Knowledge management}: The manufacturing companies' needs on the acquisition, organization, and automatic retrieval of information from different content resources, such as technical documentation, videos, images, schematics, audio, web pages and much more, in the different phases of the industrial process. 

\end{itemize} 

On the other hand, the latter category has the following subcategories: 

\begin{itemize}

    \item \textit{Product Servitization}: The innovation of the organization’s capabilities and processes to better create mutual value through a shift from selling products to selling Product-Service Systems capable of fulfilling a more comprehensive range of customer needs. 

    \item \textit{Usability improvement}: The need for solutions to user problems (e.g., the design of better user interfaces) that make it easier for the user to complete a task with effectiveness, efficiency and satisfaction by using the manufactured product. 
    
    \item \textit{Smart products}: Disruptive initiatives aimed at building a new generation of products that completely change the usage and the associated business model.
    
    \item \textit{Cost and Number of Parts or Components Reduction}: It aims at optimizing the BOM (Bill of Material) of products, thus reducing product and production management costs.
    
    \item \textit{After-Sales Service}: 
    The improvement of all those services that link the customer to the production company after selling the product (e.g., complaint management, warranty, field technical assistance responsible for installation, check-ups, out-of-warranty repairs and product disposal, usage monitoring, user analytics and profiling, automatic consumables reorder, etc.).
                
 \end{itemize} 

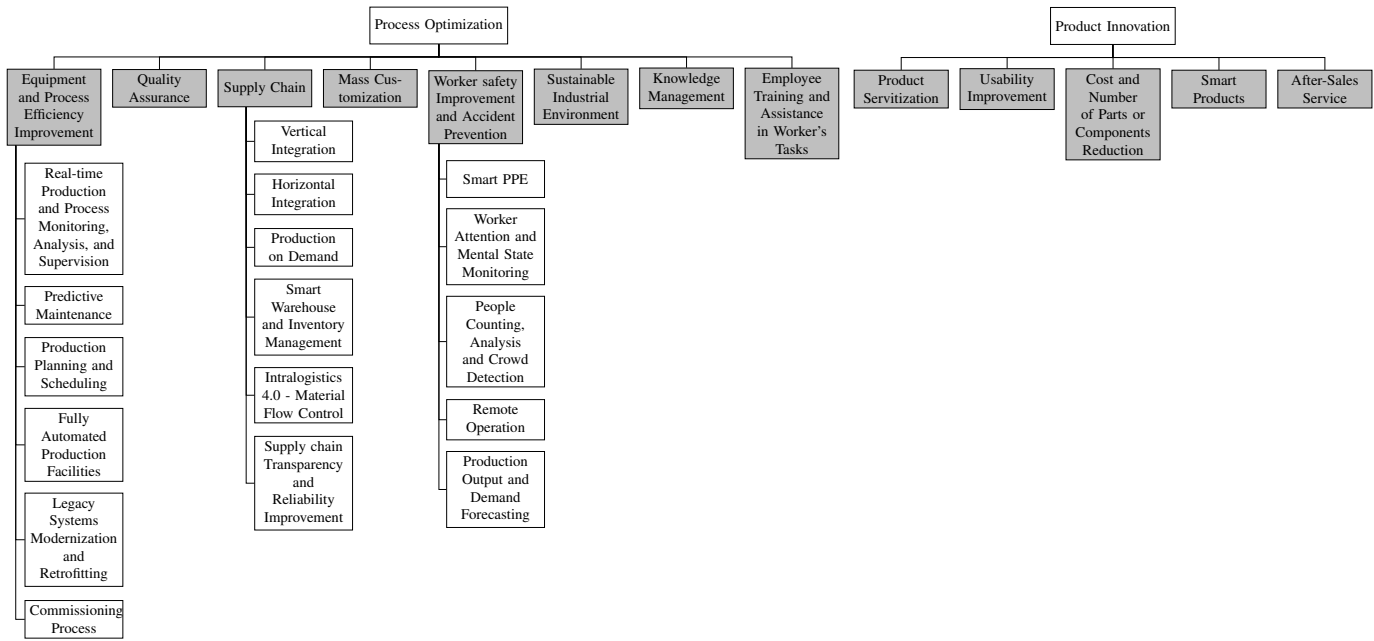
\begin {figure*}

\begin{multicols*}{2}
\vspace*{-3mm}   
\begin{adjustbox}{width=0.61\textwidth}
\begin{tikzpicture}[
criteria/.style={text centered, text width=2cm, fill=gray!50},
attribute/.style={%
    grow=down, xshift=0cm,
    text centered, text width=2.1cm,
    edge from parent path={(\tikzparentnode.225) |- (\tikzchildnode.west)}},
first/.style    ={level distance=8ex},
second/.style   ={level distance=16ex},
third/.style    ={level distance=24ex},
fourth/.style   ={level distance=32ex},
fifth/.style    ={level distance=40ex},
sixth/.style    ={level distance=40ex},
level 1/.style={sibling distance=10em}]
    \tikzstyle{every node}=[draw=black,thin,anchor=west, minimum height=2.5em]
    \node[anchor=south]{Process Optimization}
    [edge from parent fork down]

    child{node (crit1) [criteria, yshift = -0.45cm, xshift = 2cm] {Equipment and Process Efficiency Improvement}
        child[attribute,first, yshift = -1.4cm, xshift = -0.7cm]  {node {Real-time Production and Process Monitoring, Analysis, and Supervision}}
        child[attribute,second, yshift = -2.2cm, xshift = -0.7cm] {node {Predictive Maintenance}}
        child[attribute,third, yshift = -2.4cm, xshift = -0.7cm]  {node {Production Planning and Scheduling}}
        child[attribute,fourth, yshift = -3cm, xshift = -0.7cm] {node {Fully Automated Production Facilities}}
        child[attribute,fifth, yshift = -4cm, xshift = -0.7cm]  {node {Legacy Systems Modernization and Retrofitting}}
        child[attribute,sixth, yshift = -5.9cm, xshift = -0.7cm]  {node {Commissioning Process}}}
    child{node [criteria, xshift = 1cm] {Quality Assurance}}
    child{node [criteria] {Supply Chain}
        child[attribute,first, xshift = -0.25cm]  {node {Vertical Integration}}
        child[attribute,second, xshift = -0.25cm] {node {Horizontal Integration}}     
        child[attribute,third, xshift = -0.25cm]  {node {Production on Demand}}
        child[attribute,fourth, yshift = -0.4cm, xshift = -0.25cm]  {node {Smart Warehouse and Inventory Management}}
        child[attribute,fifth, yshift = -1cm, xshift = -0.25cm]  {node {Intralogistics 4.0 - Material Flow Control}}
        child[attribute,sixth, yshift = -3.1cm, xshift = -0.25cm]  {node {Supply chain Transparency and Reliability Improvement}}}
    child{node [criteria, xshift = -1cm] {Mass Customization}}
        child{node [criteria, yshift = -0.45cm, xshift = -2cm] {Worker safety Improvement and Accident Prevention}
        child[attribute,first, yshift = -0.45cm, xshift = -0.7cm]  {node {Smart PPE}}
        child[attribute,second, yshift = -0.80cm, xshift = -0.7cm] {node {Worker Attention and Mental State Monitoring}}
        child[attribute,third, yshift = -1.8cm, xshift = -0.7cm]  {node {People Counting, Analysis and Crowd Detection}}
        child[attribute,fourth, yshift = -2.4cm, xshift = -0.7cm] {node {Remote Operation}}
        child[attribute,fifth, yshift = -2.8cm, xshift = -0.7cm] {node {Production Output and Demand Forecasting}}}
    child{node [criteria, yshift = -0.20cm, xshift = -3cm] {Sustainable Industrial Environment}}
    child{node [criteria, xshift = -4cm] {Knowledge Management}}
    child{node [criteria, yshift = -0.60cm, xshift = -5cm] {Employee Training and Assistance in Worker’s Tasks}};
\end{tikzpicture}
\end{adjustbox}
\hspace*{1.9cm}
\begin{adjustbox}{width=0.38\textwidth, valign=t}
\begin{tikzpicture}[
criteria/.style={text centered, text width=2cm, fill=gray!50},
attribute/.style={%
    grow=down, xshift=0cm,
    text centered, text width=2.1cm,
    edge from parent path={(\tikzparentnode.225) |- (\tikzchildnode.west)}},
first/.style    ={level distance=8ex},
second/.style   ={level distance=16ex},
third/.style    ={level distance=24ex},
fourth/.style   ={level distance=32ex},
fifth/.style    ={level distance=40ex},
sixth/.style    ={level distance=40ex},
level 1/.style={sibling distance=10em}]
    \tikzstyle{every node}=[draw=black,thin,anchor=west, minimum height=2.5em]
    \node[anchor=south]  {Product Innovation}
    [edge from parent fork down]

    child{node (criteria) [criteria, xshift=0.88cm] {Product Servitization}}
    child{node [criteria, xshift = -.12cm] {Usability Improvement}}
    child{node [criteria, yshift = -0.60cm, xshift = -1.12cm] {Cost and Number of Parts or Components Reduction}}
    child{node [criteria,xshift = -2.12cm] {Smart Products}}
    child{node [criteria, xshift = -3.12cm] {After-Sales Service}};
\end{tikzpicture}
\end{adjustbox}

\vfill
\end{multicols*}
\caption{Identified industrial needs and challenges}
\label{fig:4}
\end{figure*}

The problems of the manufacturing industry classified according to the above categories are illustrated in Fig. \ref{fig:4}.


\subsection{Overview of the enabling technologies} \label{overview of the technologies}

Following the procedure presented in \ref{theoretical framework of taxonomies and their interoperability}, the second research question (\textbf{RQ2}) was answered. The most significant difficulty did not appear in the technology identification but their cataloging and organizing so that they can be navigated and explored intuitively for educational purposes.
This process was achieved using the European Commission report on Industry 4.0 technologies as a guideline \cite{davies_industry_nodate}. The research results and the categorization of the technological solutions are shown in Table \ref{tab:3}. Precisely, nine groups of technologies were identified:

\begin{itemize}
    \item \textit{Big Data}: Technologies related to acquiring large amounts of data from various sources, their ongoing analysis, evaluation, storage, retrieval and visualization.
    
    \item \textit{Artificial Intelligence (AI)}: Methods for building computerized systems that reason, learn from historical data, and act intelligently with little or no human intervention.
    
    \item \textit{Cloud Computing}: The offering of computing services over the Internet ("the cloud"), including servers, storage, databases, networking, software, analytics, and intelligence, to provide rapid innovation, more flexible resources, and economies of scale.
    
    \item \textit{IoT and IoE}: IoT is an interconnection of various smart devices that interact with each other and the external environment via the internet. In contrast, Internet of Everything (IoE) extends IoT, emphasizes machine-to-machine communication, and describes a more complex system that includes people and their processes.
    
    \item \textit{Digital Twins}: The virtualization of a physical object or process to analyze and simulate its behavioral model.
    
    \item \textit{Industrial Robotics}: The technologies that allow the robot to be prepared to perform production tasks and then run smoothly.
    
    \item \textit{Augmented Reality (AR) and Virtual Reality (VR)}: Technologies that work with the simulation and augmentation of the real environment.
    
    \item \textit{Additive manufacturing}: The technique of constructing a 3D object one layer - by - layer.
    
    \item \textit{Cybersecurity Technologies}: Technologies to defend against cyber attacks on systems, networks, programs, devices, and data.

\end{itemize} 

Table \ref{tab:3} hows the I4.0 enabling technologies grouped into specific categories that underpin the industrial needs. The whole taxonomy, the studied literature and the definitions of all terms are given in \cite{riccardo_amadio_2022_7215558}.

\subsection{Website presentation of the taxonomy} \label{website presentation}
The above taxonomy can provide a valuable way to link the two studied domains by presenting a sufficient number of selected references for each term. However, a web tool through which the interested party can explore the taxonomy and its references was built to provide a shared knowledge tool for better implementation of I4.0 (\textbf{RQ3}) and make it more usable to the intended users. This website, named "Planet4 Taxonomy Explorer", is available on the internet at the following link: \url{http://taxonomy.planet4project.eu/}. 

The web application development began with the taxonomy construction, and the current version is already in its second iteration. The website provides fundamental means of exploration, based mainly on a search bar, a tree view presenting the structure of the taxonomy and a results page. The search engine that allows the search bar to work uses the TagMe API \cite{10.1145/1871437.1871689} to identify user input by comparing it first with the Wikipedia database and then with our I4.0 taxonomy. This way, the most relevant results -i.e., the academic and grey sources that share the fundamental terms listed within the taxonomy- associated with the query can be viewed on a dedicated results page. On the same page, each result (or article) has its taxonomy terms presented as keywords. Thus, users can quickly identify significant articles -and filter them depending on their business macro need or technology- to find solutions or deepen their knowledge. In \cite{riccardo_amadio_2022_7215558}, the whole repository of the search engine, together with the database of the taxonomy, is provided.

\subsection{Taxonomy application in problem-solving} \label{taxonomy application}

To solve industrial problems using the taxonomy, the user can search for the term that best reflects the business challenge and discover the publications in which the same or similar problems are addressed and the technologies mentioned in them. Information about the technologies needed can help assess whether the company has the necessary resources to solve the problem on its own, whether external help is required and to what extent. If the company has the appropriate epistemological background but inadequate experience, the company specialists can study the indicated publications, which may help them solve problems independently.

The first example of applying the taxonomy presents a problem with production flow monitoring. The challenge was identified by a furniture manufacturer operating in Poland, which produces furniture fronts and complete furniture systems and supplies its products to the Polish, Czech, Slovak, and Ukrainian markets. The company wanted: to monitor a) the production flow of all types of furniture fronts, b) the order status, c) the location of each production batch, and d) the time remaining until the end of the production of a given batch.

Analyzing the challenge description and the industrial problems and needs classified in the taxonomy, it follows that the challenge matches the subcategory "Real-time Production and Process Monitoring, Analysis, and Supervision". During the review of the sources indicated by the taxonomy for solving this problem, 12 publications related to RFID tags and readers were identified. In turn, 12 sources stated the possibility of using RTOS in the discussed problem. Five publications indicate the need for Time Series Databases. At the same time, forty references propose Machine Learning methods to predict the remaining time to complete a production order. Deep Learning techniques are the most widely used in such problems (13 sources), followed by Supervised Learning techniques (10 sources). The last element of the challenge solution proposal is Grafana, which was indicated in 2 sources. Therefore, the following solution was proposed as a starting point:

\begin{itemize}
    \setlength{\parskip}{0pt}
    \setlength{\itemsep}{0pt plus 1pt}
    \item Edge device:
        \begin{itemize}
        \setlength{\parskip}{0pt}
        \setlength{\itemsep}{0pt plus 1pt}
                \item Attach RFID tags to containers or products for capturing information about their status and location. RFID readers should be located at the indicated points in the process.
                \item Use the real-time operating system (RTOS) to maintain batch location information and enable real-time alert systems during delays.
        \end{itemize}            
    \item Cloud service that:
        \begin{itemize}
        \setlength{\parskip}{0pt}
        \setlength{\itemsep}{0pt plus 1pt}
            \item Uses the Time Series Database (TSDB) to store the incoming data.
            \item Runs Deep Learning models that predict the time remaining to complete a production order.
            \item Sends this data to the Grafana engine to visualize the data in dashboards for the end-user.
        \end{itemize}    
\end{itemize}

\begin{figure*}[htbp]
  \centering
    \includegraphics[width=0.7\textwidth]{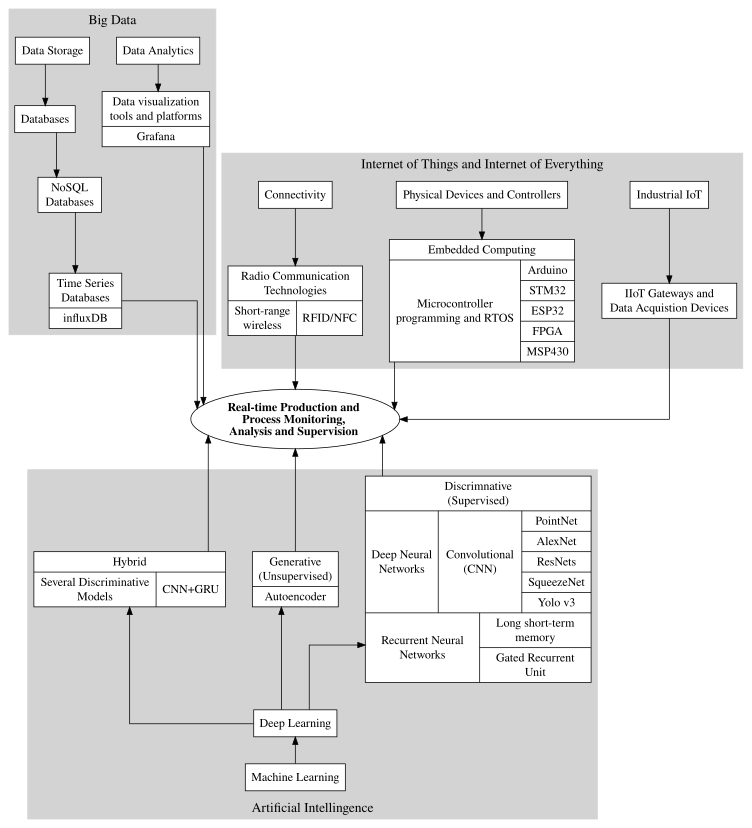}
  \caption{Fragment of the taxonomy with the solution of the production flow monitoring problem}
  \label{fig:6}
\end{figure*}

Fig. \ref{fig:6} presents a fragment of the taxonomy that includes the described proposal to solve the problem of production flow monitoring.


The second example considers a problem related to predictive maintenance. The challenge was proposed by a Polish company that is a pioneer in the production of fasteners. The company would like to implement a system that prevents machine failures or minimizes the number and duration of failures. The company has sixty modern presses monitored during production to ensure process stability.

Analyzing the challenge description and the industrial problems and needs classified in the taxonomy, it follows that the challenge matches the subcategory "Predictive Maintenance" (Fig. \ref{fig:4}). The taxonomy exploration revealed 16 publications that use sensors and Industrial Communication Protocols for Predictive Maintenance. Over the above, 10 publications concerned Microcontroller programming and RTOS employment, while Time Series Databases were recommended in 13 publications. At the same time, 67 sources present Machine Learning approaches in such tasks. Supervised Learning is the most widespread Machine Learning task, with 18 references, followed by Deep Learning, with 11 references. Finally, two sources indicate the need to use Grafana in this problem. Hence, the following solution was proposed as a starting point:

\begin{itemize}
    \setlength{\parskip}{0pt}
    \setlength{\itemsep}{0pt plus 1pt}
    \item Edge device:
        \begin{itemize}
        \setlength{\parskip}{0pt}
        \setlength{\itemsep}{0pt plus 1pt}
                \item Identify sensors already embedded in machines for data extraction, and connect additional sensors where needed. Then, communicate with the machines’ control system using industrial communication protocols.
                \item Use a real-time operating system (RTOS) to collect and process data on measured parameter values over time and enable real-time alert systems.
        \end{itemize}            
    \item Cloud / Edge service that:
        \begin{itemize}
        \setlength{\parskip}{0pt}
        \setlength{\itemsep}{0pt plus 1pt}
            \item Uses the Time Series Database (TSDB) to store the incoming data.
            \item Runs Supervised Learning models that predict the possibility of a failure and will indicate the predicted failure location and proposed actions.
            \item Sends this data to the Grafana engine to visualize the data in dashboards for the end-user.
        \end{itemize}    
\end{itemize}

\begin{figure*}[htbp]
  \centering
    \includegraphics[width=1\textwidth]{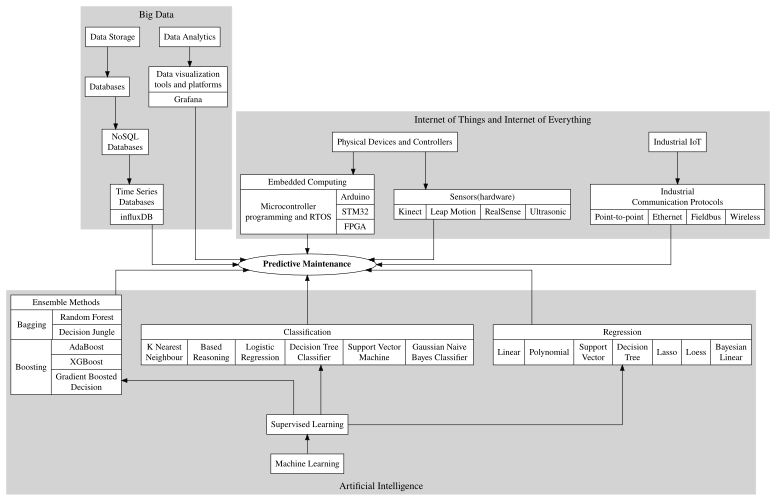}
  \caption{Fragment of the taxonomy with the solution of the predictive maintenance problem}
  \label{fig:7}
\end{figure*}

Fig. \ref{fig:7} shows a fragment of the taxonomy containing the described proposal to solve the problem of predictive maintenance.


\begin{table*}
\centering
\caption{List of I4.0 enabling technologies.}
\hspace*{-3cm}\resizebox{0.84\textwidth}{!}{%
\begin{tabular}{p{0.352\textwidth}p{0.352\textwidth}p{0.28\textwidth}} 
\begin{tabular}{|l|}
\hline
\multicolumn{1}{|c|}{\textbf{Big Data}} \\
\hline
\multicolumn{1}{|l|}{\begin{tabular}{@{}l@{}}{\labelitemi}\hspace{\dimexpr\labelsep+0.5\tabcolsep}Big Data Frameworks\\{\labelitemi}\hspace{\dimexpr\labelsep+0.5\tabcolsep}Data Sources/Ingestion\\\hspace{0.5\leftmargin}{\labelitemii}\hspace{\dimexpr\labelsep+0.5\tabcolsep}Streaming and Messaging\\\hspace{0.5\leftmargin}{\labelitemii}\hspace{\dimexpr\labelsep+0.5\tabcolsep}Orchestration and Pipelines\\\hspace{0.5\leftmargin}{\labelitemii}\hspace{\dimexpr\labelsep+0.5\tabcolsep}Query/Data Flow\\{\labelitemi}\hspace{\dimexpr\labelsep+0.5\tabcolsep}Data Storage\\\hspace{0.5\leftmargin}{\labelitemii}\hspace{\dimexpr\labelsep+0.5\tabcolsep}Databases\\\hspace{0.5\leftmargin}{\labelitemii}\hspace{\dimexpr\labelsep+0.5\tabcolsep}Data Warehouses\\{\labelitemi}\hspace{\dimexpr\labelsep+0.5\tabcolsep}Data Analytics\\\hspace{0.5\leftmargin}{\labelitemii}\hspace{\dimexpr\labelsep+0.5\tabcolsep}Unified Data Analytics Engines\\\hspace{0.5\leftmargin}{\labelitemii}\hspace{\dimexpr\labelsep+0.5\tabcolsep}Unified stream-processing and batch-processing frameworks\hspace*{1.2mm}\\\hspace{0.5\leftmargin}{\labelitemii}\hspace{\dimexpr\labelsep+0.5\tabcolsep}Business Intelligence (BI) Tools\\\hspace{0.5\leftmargin}{\labelitemii}\hspace{\dimexpr\labelsep+0.5\tabcolsep}Data Visualization Tools and Platforms\\\hspace{0.5\leftmargin}{\labelitemii}\hspace{\dimexpr\labelsep+0.5\tabcolsep}Logging and Monitoring\\\hspace{0.5\leftmargin}{\labelitemii}\hspace{\dimexpr\labelsep+0.5\tabcolsep}Spreadsheet Applications\\\hspace{0.5\leftmargin}{\labelitemii}\hspace{\dimexpr\labelsep+0.5\tabcolsep}Data Mining\\\hspace{0.5\leftmargin}{\labelitemii}\hspace{\dimexpr\labelsep+0.5\tabcolsep}Process Mining\end{tabular}}\\
\hline
\end{tabular}
&
\hspace*{1.65cm}\begin{tabular}{|l|}
\hline
\multicolumn{1}{|c|}{\textbf{Artificial Intelligence}} \\
\hline
\multicolumn{1}{|l|}{\begin{tabular}{@{}l@{}}{\labelitemi}\hspace{\dimexpr\labelsep+0.5\tabcolsep}Machine Learning\\\hspace{0.5\leftmargin}{\labelitemii}\hspace{\dimexpr\labelsep+0.5\tabcolsep}Supervised Learning\\\hspace{0.5\leftmargin}{\labelitemii}\hspace{\dimexpr\labelsep+0.5\tabcolsep}Unsupervised Learning\\\hspace{0.5\leftmargin}{\labelitemii}\hspace{\dimexpr\labelsep+0.5\tabcolsep}Deep Learning\\\hspace{0.5\leftmargin}{\labelitemii}\hspace{\dimexpr\labelsep+0.5\tabcolsep}Transfer Learning\\\hspace{0.5\leftmargin}{\labelitemii}\hspace{\dimexpr\labelsep+0.5\tabcolsep}Reinforcement Learning\\\hspace{0.5\leftmargin}{\labelitemii}\hspace{\dimexpr\labelsep+0.5\tabcolsep}Deep Reinforcement Learning\\\hspace{0.5\leftmargin}{\labelitemii}\hspace{\dimexpr\labelsep+0.5\tabcolsep}Semi-Supervised Learning\\\hspace{0.5\leftmargin}{\labelitemii}\hspace{\dimexpr\labelsep+0.5\tabcolsep}Federated learning\\{\labelitemi}\hspace{\dimexpr\labelsep+0.5\tabcolsep}Computer Vision\\{\labelitemi}\hspace{\dimexpr\labelsep+0.5\tabcolsep}Natural Language Processing, Natural Language Generation\\{\labelitemi}\hspace{\dimexpr\labelsep+0.5\tabcolsep}Intelligent Agents and Multiagent Systems\\{\labelitemi}\hspace{\dimexpr\labelsep+0.5\tabcolsep}Soft Computing\\\hspace{0.5\leftmargin}{\labelitemii}\hspace{\dimexpr\labelsep+0.5\tabcolsep}Fuzzy Set Theory\\\hspace{0.5\leftmargin}{\labelitemii}\hspace{\dimexpr\labelsep+0.5\tabcolsep}Neurocomputing\\\hspace{0.5\leftmargin}{\labelitemii}\hspace{\dimexpr\labelsep+0.5\tabcolsep}Optimization Techniques\\\hspace{0.5\leftmargin}{\labelitemii}\hspace{\dimexpr\labelsep+0.5\tabcolsep}Probabilistic Reasoning\end{tabular}}\\
\hline
\end{tabular}
&
\hspace{2.8cm}\begin{tabular}{|l|}
\hline
\multicolumn{1}{|c|}{\textbf{Cloud Computing}} \\
\hline
\multicolumn{1}{|l|}{\begin{tabular}{@{}l@{}}{\labelitemi}\hspace{\dimexpr\labelsep+0.5\tabcolsep}Infrastructure as a Service (IaaS)\\\hspace{0.5\leftmargin}{\labelitemii}\hspace{\dimexpr\labelsep+0.5\tabcolsep}Cloud Data Storage and Computing\\{\labelitemi}\hspace{\dimexpr\labelsep+0.5\tabcolsep}Platform as a Service (PaaS)\\\hspace{0.5\leftmargin}{\labelitemii}\hspace{\dimexpr\labelsep+0.5\tabcolsep}Device Management\\\hspace{0.5\leftmargin}{\labelitemii}\hspace{\dimexpr\labelsep+0.5\tabcolsep}Operating System\\{\labelitemi}\hspace{\dimexpr\labelsep+0.5\tabcolsep}Software as a Service (SaaS)\\\hspace{0.5\leftmargin}{\labelitemii}\hspace{\dimexpr\labelsep+0.5\tabcolsep}Media Streaming Software Platforms\\\hspace{0.5\leftmargin}{\labelitemii}\hspace{\dimexpr\labelsep+0.5\tabcolsep}Website Building\\\hspace{0.5\leftmargin}{\labelitemii}\hspace{\dimexpr\labelsep+0.5\tabcolsep}IoT Analytics Software and Platforms\\{\labelitemi}\hspace{\dimexpr\labelsep+0.5\tabcolsep}Infrastructure as Code (IaC)\\\hspace{0.5\leftmargin}{\labelitemii}\hspace{\dimexpr\labelsep+0.5\tabcolsep}Provisioning Tools\\{\labelitemi}\hspace{\dimexpr\labelsep+0.5\tabcolsep}Container Technology (Container as a Service)\\\hspace{0.5\leftmargin}{\labelitemii}\hspace{\dimexpr\labelsep+0.5\tabcolsep}Containerization Platform\\\hspace{0.5\leftmargin}{\labelitemii}\hspace{\dimexpr\labelsep+0.5\tabcolsep}Container Orchestration\\{\labelitemi}\hspace{\dimexpr\labelsep+0.5\tabcolsep}Serverless Programming\\{\labelitemi}\hspace{\dimexpr\labelsep+0.5\tabcolsep}Edge Computing\\{\labelitemi}\hspace{\dimexpr\labelsep+0.5\tabcolsep}Fog Computing\end{tabular}}\\
\hline
\end{tabular}
\\[28.5mm]


\begin{tabular}{|l@{\hspace*{3cm}}|}
\hline
\multicolumn{1}{|c|}{\textbf{IoT and IoE}} \\
\hline
\multicolumn{1}{|l|}{\begin{tabular}{@{}l@{}}{\labelitemi}\hspace{\dimexpr\labelsep+0.5\tabcolsep}Industrial IoT\\\hspace{0.5\leftmargin}{\labelitemii}\hspace{\dimexpr\labelsep+0.5\tabcolsep}Industrial Communication Protocols\\\hspace{0.5\leftmargin}{\labelitemii}\hspace{\dimexpr\labelsep+0.5\tabcolsep}Industrial (IoT) Gateways and Data Acquisition Devices\\\hspace{0.5\leftmargin}{\labelitemii}\hspace{\dimexpr\labelsep+0.5\tabcolsep}Software Data Adapters
\\{\labelitemi}\hspace{\dimexpr\labelsep+0.5\tabcolsep}Physical Devices and Controllers\\\hspace{0.5\leftmargin}{\labelitemii}\hspace{\dimexpr\labelsep+0.5\tabcolsep}Embedded Computing\\\hspace{0.5\leftmargin}{\labelitemii}\hspace{\dimexpr\labelsep+0.5\tabcolsep}Sensors (hardware)\\{\labelitemi}\hspace{\dimexpr\labelsep+0.5\tabcolsep}Signal Processing\\{\labelitemi}\hspace{\dimexpr\labelsep+0.5\tabcolsep}Connectivity\\\hspace{0.5\leftmargin}{\labelitemii}\hspace{\dimexpr\labelsep+0.5\tabcolsep}Radio Communication Technologies\\\hspace{0.5\leftmargin}{\labelitemii}\hspace{\dimexpr\labelsep+0.5\tabcolsep}Optical Communication Technologies\\\hspace{0.5\leftmargin}{\labelitemii}\hspace{\dimexpr\labelsep+0.5\tabcolsep}IoT Messaging Protocols\\\hspace{0.5\leftmargin}{\labelitemii}\hspace{\dimexpr\labelsep+0.5\tabcolsep}Application Programming Interfaces and Programming Tools\\{\labelitemi}\hspace{\dimexpr\labelsep+0.5\tabcolsep}IoE (Internet of Everything)\end{tabular}}\\
\hline
\end{tabular}

&
\hspace*{1.65cm}\begin{tabular}{|l|}
\hline
\multicolumn{1}{|c|}{\textbf{Digital Twins}} \\
\hline\\[1cm]
\multicolumn{1}{|l|}{\begin{tabular}{@{\labelitemi\hspace{\dimexpr\labelsep+0.5\tabcolsep}}l@{}}Computer-aided design (CAD) Software\\Finite Element Analysis (FEA) Software\\Simulation Software\\DTs Management and Orchestration Frameworks\hspace*{1.25cm}\\Digital Twin Data Modelling\\Virtual Process Controllers (VPC)\end{tabular}}   \\[2cm]
\hline
\end{tabular}

&

\hspace*{2.8cm}\begin{tabular}{|l|}
\hline
\multicolumn{1}{|c|}{\textbf{Industrial Robotics}} \\
\hline\\[1.63cm]
\multicolumn{1}{|l|}{\begin{tabular}{@{\labelitemi\hspace{\dimexpr\labelsep+0.5\tabcolsep}}l@{}}Offline Programming and Simulation\hspace*{1.15cm}\\Middleware\end{tabular}}\\[2cm]
\hline
\end{tabular}
\\[-1.5mm]

\begin{tabular}{|l|}
\hline
\multicolumn{1}{|c|}{\textbf{Augmented Reality (AR) and Virtual Reality (VR)}} \\
\hline\\[1.56cm]
\multicolumn{1}{|l|}{\begin{tabular}{@{}l@{}}{\labelitemi}\hspace{\dimexpr\labelsep+0.5\tabcolsep}VR\\\hspace{0.5\leftmargin}{\labelitemii}\hspace{\dimexpr\labelsep+0.5\tabcolsep}VR glasses\\{\labelitemi}\hspace{\dimexpr\labelsep+0.5\tabcolsep}AR\\\hspace{0.5\leftmargin}{\labelitemii}\hspace{\dimexpr\labelsep+0.5\tabcolsep}AR glasses\\\hspace{0.5\leftmargin}{\labelitemii}\hspace{\dimexpr\labelsep+0.5\tabcolsep}AR Software Development Kits\\{\labelitemi}\hspace{\dimexpr\labelsep+0.5\tabcolsep}AR and VR Software development, Platforms and Technologies\end{tabular}} \\[3cm]
\hline
\end{tabular}

&
\hspace*{1.65cm}\begin{tabular}{|l|}
\hline
\multicolumn{1}{|c|}{\textbf{Additive Manufacturing}} \\
\hline\\[2.2cm]
\multicolumn{1}{|l|}{\begin{tabular}{@{\labelitemi\hspace{\dimexpr\labelsep+0.5\tabcolsep}}l@{}}3D Printers \hspace*{5.5cm}\\3D Printing Technologies\end{tabular}} \\[3cm]
\hline
\end{tabular}
&
\hspace*{2.8cm}\begin{tabular}{|l|}
\hline
\multicolumn{1}{|c|}{\textbf{Cybersecurity Technologies}} \\
\hline
\multicolumn{1}{|l|}{\begin{tabular}{@{}l@{}}{\labelitemi}\hspace{\dimexpr\labelsep+0.5\tabcolsep}Security Virtualization\\\hspace{0.5\leftmargin}{\labelitemii}\hspace{\dimexpr\labelsep+0.5\tabcolsep}Virtual Machine Monitor (VMM)\\{\labelitemi}\hspace{\dimexpr\labelsep+0.5\tabcolsep}Data Protection\\\hspace{0.5\leftmargin}{\labelitemii}\hspace{\dimexpr\labelsep+0.5\tabcolsep}Secure Communication Protocols\\\hspace{0.5\leftmargin}{\labelitemii}\hspace{\dimexpr\labelsep+0.5\tabcolsep}Key Management System (KMS)\hspace*{1.22cm}\\\hspace{0.5\leftmargin}{\labelitemii}\hspace{\dimexpr\labelsep+0.5\tabcolsep}Public Key Infrastructure (PKI)\\\hspace{0.5\leftmargin}{\labelitemii}\hspace{\dimexpr\labelsep+0.5\tabcolsep}Encryption\\\hspace{0.5\leftmargin}{\labelitemii}\hspace{\dimexpr\labelsep+0.5\tabcolsep}Tokenization\\\hspace{0.5\leftmargin}{\labelitemii}\hspace{\dimexpr\labelsep+0.5\tabcolsep}Blockchain\\{\labelitemi}\hspace{\dimexpr\labelsep+0.5\tabcolsep}Identity and Access Management\\\hspace{0.5\leftmargin}{\labelitemii}\hspace{\dimexpr\labelsep+0.5\tabcolsep}Protocols\\\hspace{0.5\leftmargin}{\labelitemii}\hspace{\dimexpr\labelsep+0.5\tabcolsep}User Management\\\hspace{0.5\leftmargin}{\labelitemii}\hspace{\dimexpr\labelsep+0.5\tabcolsep}Authentication\\\hspace{0.5\leftmargin}{\labelitemii}\hspace{\dimexpr\labelsep+0.5\tabcolsep}Authorization\\{\labelitemi}\hspace{\dimexpr\labelsep+0.5\tabcolsep}Security Operations\\\hspace{0.5\leftmargin}{\labelitemii}\hspace{\dimexpr\labelsep+0.5\tabcolsep}Change Management\\\hspace{0.5\leftmargin}{\labelitemii}\hspace{\dimexpr\labelsep+0.5\tabcolsep}Threat Detection and Analysis\\{\labelitemi}\hspace{\dimexpr\labelsep+0.5\tabcolsep}Foundational Security\\\hspace{0.5\leftmargin}{\labelitemii}\hspace{\dimexpr\labelsep+0.5\tabcolsep}Network\end{tabular}}\\
\hline
\end{tabular}
\\
\end{tabular}
}

\label{tab:3}
\end{table*}


\section{Conclusions} \label{conclusions}

\subsection{General Conclusions} \label{general conclusions}

This paper presents the development of an enhanced tool based on a taxonomy that links industrial needs and challenges with available solutions technologies by analysing 441 academic and grey sources. This tool will help academics and industry stakeholders identify the relevant business scenarios in the context of I4.0 and design/deploy their solutions based on the latest technologies and existing works documented in the literature. 
A systematic process based on state-of-the-art methodologies has been followed for its development.  
In contrast to other works building similar tools, this paper presents a holistic view of I4.0. Moreover, thanks to this significant amount of information, several pertinent statistics related to I4.0 have also been obtained.  Finally, a user-friendly web interface has also been implemented to enhance the tool's usability, and its application in two actual use cases has been detailed to facilitate its adoption.

\subsection{Work Limitations} \label{work limitations}

The developed taxonomy shows three phases of subjectivity during its development and use by the stakeholders. The full involvement of the human factor in the taxonomy development implies that a) the taxonomic classification and conceptualization of needs and technologies and b) the manual attribution of each bibliographic source was at the authors' discretion. Therefore, it is understandable that there were difficulties in finalizing the taxonomy, mainly regarding industrial needs. A representative example was articles dealing with monitoring industrial equipment's condition, which some consider belonging to the "Real-time Production and Process Monitoring, Analysis, and Supervision" category, while others felt it belongs to the "Predictive Maintenance" category.
In the next phase, companies that pilot used the taxonomy assigned the same challenge to different taxonomy concepts due to the different ways of approaching it. For example, industries that wanted to estimate the maintenance time of a failed industrial equipment either chose "Predictive Maintenance" as the concept that best reflects their problem or "Production Planning and Scheduling". To deal with the difficulties above to some extent, we maintained the challenges hierarchy at two levels (Fig. \ref{fig:4}) without further categorization, as in the case of technologies. At the same time, the given definitions clarified the sub-challenges included in each category.

\subsection{Future Work} \label{future work}

The taxonomies presented here can be the starting point for building an I4.0 ontology, thus allowing for an even more formal and standardized representation. At the same time, it would be possible to link it to other already existing ontologies to achieve interoperability and address knowledge domains not dealt with in this study but still linked to the general topic of I4.0. Moreover, an ontology would form the basis for building the knowledge graph allowing the retrieval of information on the website and enabling new features, such as adding and indexing new content and better maintainability and updatability.

Future work should also focus on expanding the taxonomy by including more sources and updating the website's interface and capabilities to provide a more scalable and maintainable data structure offering users a more consistent and understandable presentation of its contents. One such functionality is the dynamic calculation and displays (based on the database's current status) of statistics related to the industry's current challenges, the I4.0 enabling technologies, and the analyzed sources such as "Which are the most cited articles?"; "Which are the most presented industrial needs?"; "What are the industrial problems with the least technological solutions?"; "Which are the most presented technologies?" or "Which are the most used technologies (and their combinations)?". A fragment of the above questions' statistics is shown in Fig. \ref{statistics} for the existing database's content. Another one is importing new sources (either by researchers or automatically) to maintain the knowledge base constantly up-to-date, which are in the immediate plans. We could also deal with the development of a set of taxonomy application examples that would make it easier for users to use the taxonomy for various business needs. Such examples could appear on the website as a "case studies" section. Therefore, in the effort to develop a tool that will contribute to the I4.0 vision realization from the industry and academia perspective, any contributions from the scientific community are welcome.

\bibliographystyle{IEEEtran}
\bibliography{mybibfile}

\end{document}